\documentclass[twocolumn]{aastex61}

\usepackage{amsmath}
\usepackage{bm}
\usepackage{lipsum}

\newcommand{\pd}[2]{\ensuremath{\frac{\partial{#1}}{\partial{#2}}}}

\newcommand{\cmmnt}[1]{\ignorespaces}

\begin{document}

\title{On the geometry of curvature radiation and implications for subpulse drifting}

\author{S. J. McSweeney}
\author{N. D. R. Bhat}
\author{S. E. Tremblay}
\affiliation{International Centre for Radio Astronomy Research (ICRAR), Curtin University, 1 Turner Ave., Technology Park, Bentley, 6102, W.A., Australia}
\affiliation{ARC Centre of Excellence for All-sky Astronomy (CAASTRO)}
\author{A. A. Deshpande}
\affiliation{Raman Research Institute (RRI), C. V. Raman Avenue, Sadashivanagar, Bengaluru---560 080, India}
\author{G. Wright}
\affiliation{Jodrell Bank Centre for Astrophysics, School of Physics and Astronomy, University of Manchester, M13 9PL, UK}

\begin{abstract}
    The phenomenon of subpulse drifting offers unique insights into the emission geometry of pulsars, and is commonly interpreted in terms of a rotating carousel of ``spark'' events near the stellar surface.
    We develop a detailed geometric model for the emission columns above a carousel of sparks that is entirely calculated in the observer's inertial frame, and which is consistent with the well-understood rotational effects of aberration and retardation.
    We explore the observational consequences of the model, including (1) the appearance of the reconstructed beam pattern via the cartographic transform and (2) the morphology of drift bands and how they might evolve as a function of frequency.
    The model, which is implemented in the software package PSRGEOM, is applicable to a wide range of viewing geometries, and we illustrate its implications using PSRs B0809+74 and B2034+19 as examples.
    Some specific predictions are made with respect to the difference between subpulse evolution and microstructure evolution, which provides a way to further test our model.
\end{abstract}


\section{Introduction}

The phenomenon of subpulse drifting \citep{Drake1968,Backer1973} offers unique insights into the emission physics of pulsars, if the correct interpretation can be found.
One popular interpretation has its basis in the carousel model, originally proposed by \citet{Ruderman1975}, in which the geometry of the emitted pulsar beam is intimately connected with the geometry of a circular pattern of sparks near the stellar surface that rotate around the magnetic pole.
More recently, \citet{Deshpande1999,Deshpande2001} developed a technique for mapping the two-dimensional emission beam that overcomes the inherent difficulty arising from the fact that a fixed observer's line of sight only ever makes a one-dimensional cut through it.
This ``cartographic transform'' technique requires sufficiently long and stable pulse sequences\footnote{Pulse sequences can be interrupted by nulls and mode changes on quasi-random time scales, whose statistics are different from pulsar to pulsar.} in order to ``fill in'' the map completely enough to see the beam's global structure.

The cartographic transform involves a number of parameters, including $\alpha$, the angle between the rotation and magnetic axes, $\zeta$, the angle between the rotation axis and the line of sight, $N$, the number of beamlets in the carousel, and $P_4$, the rotation time of the carousel.
If these parameters are either not known beforehand, or are coarsely estimated by other means, the cartographic transform can be used to test the viability of sets of parameters.
In general, if one or more of the parameters put into the transform are slightly incorrect, the resulting polar beam pattern will be smeared and hence would be relatively devoid of structure.
Then, a pulse sequence reconstructed from it will not correlate well with the original pulse sequence.
However, the test is not entirely sufficient for finding the correct set of parameters as degeneracies may exist (e.g. due to aliasing of the rotating spark pattern modulo the pulsar rotation), which would have to be resolved by other means.

In the case of PSR B0809+74, \citet{Rankin2006} found that frequent mode changes, nulls, and ``memory'' of drift band phase across the nulls, all conspired to make it difficult to distinguish between a handful of solutions that all produced credible beam emission maps (see their paper, and references therein, for a full account of the difficulties encountered in their analysis).
One of their solutions ($\alpha \approx 9^\circ$, $\zeta \approx 13.5^\circ$, $N = 10$) produced a polarised beam map with a striking asymmetry, where the beamlets of one of the orthogonal polarisations appear to be skewed in a common azimuthal direction (see their Fig. 4).

Under the assumption that the geometric parameters are correct, the authors connect this asymmetrical feature to ``absorption'' \citep{Rankin2006a}, which is an empirical phenomenon, currently lacking physical justification.
Here, we ask whether this kind of beam asymmetry can possibly arise from purely geometric considerations once rotational effects are taken into account.
In this context, the primary relevant rotational effects are aberration and retardation (hereafter, AR effects), which are both height-dependent effects that cause emission to appear at an earlier phase than otherwise expected \citep{Blaskiewicz1991,Dyks2003}.

The treatment of aberration begins by making a few simple assumptions about how the emitted radiation would appear in the corotating frame (CF), and then using the principles of relativistic aberration to translate the Poynting vector into the observer's inertial frame (IF).
\citet{Dyks2003} show that if you assume that the Poynting vector is parallel to the tangent of the local magnetic field at the emission point, then the angle between the Poynting vector and the tangent of the local magnetic field in the IF, $\eta_\text{ab}$, is, to first order,
\begin{equation}
    \eta_\text{ab} \approx r^\prime\sin\theta_z,
    \label{eqn:dyks_eta}
\end{equation}
where $r^\prime = r/r_L$, $r$ is the distance of the emission point from the origin placed at the stellar centre, $r_L = c/\Omega$ is the light cylinder radius, $\Omega = 2\pi/P$ is the angular frequency of pulsar rotation, and $\theta_z$ is the angle that the position vector of the emission point makes with the rotation axis $\hat{\bm{\Omega}} \equiv \hat{\bm{z}}$.
The observable effect of this aberrational correction is a shift of the profile towards earlier phases,
\begin{equation}
    \Delta\phi_\text{ab} \approx \frac{\eta_\text{ab}}{\sin\zeta}.
\end{equation}
\citet{Dyks2003} point out, however, that the assumption of parallel vectors in the CF requires that the Lorentz factors of the emitting particles are relatively high, a requirement that is necessary in any case for curvature radiation to be in the observed range of radio frequencies.

The necessity of translating between the CF and the IF would not be required if the trajectories of the particles were already known in the IF.
In this case, the Poynting vector of the curvature radiation could simply be identified with the instantaneous velocity of the particles (and of course across the width of the individual particle beam, which goes as $\sim 1/\gamma$).
It is well established that the magnetic field lines are ``frozen into'' the corotating magnetosphere, and that particles are tightly constrained to move along magnetic field lines (at least, well within the light cylinder radius) as ``beads on a wire'' \citep[e.g.][]{Goldreich1969,Sturrock1971,Ruderman1975}.

The problem of determining the trajectory of a particular particle on a particular field line from first principles is a non-trivial exercise.
In \citet{Thomas2007}, the equations of motion for a particle in the (vacuous) magnetosphere are derived and evaluated numerically to produce particle trajectories that depend on both the functional form of the magnetic field and the magnetic inclination angle $\alpha$.
They found that there is significant contribution, due to rotation effects, to the curvature of field lines on the leading side of the magnetic axis as compared to those on the trailing side.
Indeed, even particles on field lines very close to the magnetic axis have trajectories with significant rotation-induced curvature even though the field lines themselves have very small curvature.

\citet{Thomas2007} remind us that in the emission region, the various forces acting on a magnetospheric particle are expected to balance in such a way that the particle is tightly constrained to remain on a single field line.
This is because, in the words of \citet{Radhakrishnan2001}, ``any transverse momentum and energy would be radiated away `instantly', and the charged particles would be in their lowest Landau levels.''
This implies that in regions where this balance is maintained, there is a relatively simple way to deduce the velocities of particles in the emission region by geometric arguments, without needing to evaluate the competing forces acting on the particles.
Indeed, a sufficient set of assumptions to derive a particle's trajectory are that (1) we know the time-dependent magnetic field in the IF (and that the observer is at rest with respect to the centre of the star), (2) particles (in the emission region) move one-dimensionally along field lines as beads on a wire, and (3) we know the particle's speed in the IF, $\beta \equiv v/c$.
In Section \S\ref{sec:vel_derivation} we derive the particle velocities under the above assumptions, using a simple dipole field, and compare the results of this derivation to the more traditional approach via aberration as presented by \citet{Dyks2003} and others.
We then proceed in Section \S\ref{sec:beam_shape} to use the derived analytical expression of the velocity field to predict the shape of the beam, both analytically in the case of an aligned rotator, and numerically in the general case.
The geometric model is then applied in Section \S\ref{sec:applications} to PSRs B0809+74 and B2034+19, two subpulse drifters with contrasting drift band morphologies, to demonstrate the viability of the geometric approach.
Finally, we include a discussion on the conditions required to observe various geometric effects in drift bands.

\section{Derivation of the velocity and acceleration fields}
\label{sec:vel_derivation}

\begin{figure}[t]
    \centering
    \includegraphics[scale=0.4]{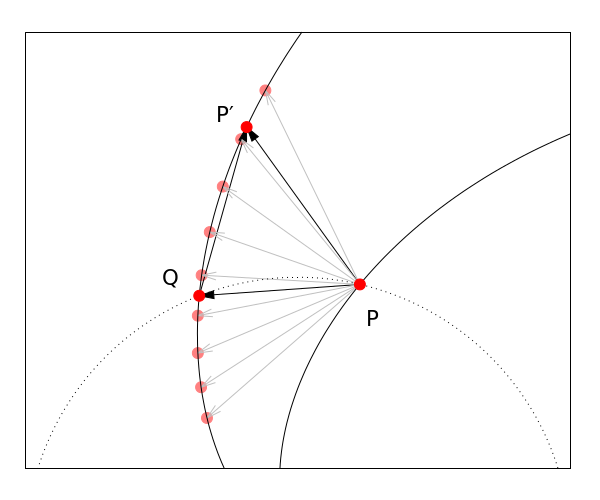}
    \caption{A schematic of the possible trajectories of a particle corotating with a magnetic field line. In the absence of any radial motion, a particle located at $P$ at time $t$ would end up at $Q$ at time $t + \Delta t$. In general, the particle may move in a limited range of directions, as illustrated by the grey arrows, each corresponding to a specific speed measured by an observer in the IF. One arbitrary direction in this range is indicated by the point $P^\prime$ and the black arrows.}
    \label{fig:beadonwire}
\end{figure}

The requirement of particles to stay on a corotating field line provides a strong geometric constraint on the possible trajectories of the particle.
Let $\vec{C}(s,t)$ be a parametrisation of a given field line.
Then corotation implies that $\vec{C}(s,t+\Delta t) = R_z(\Delta t)\vec{C}(s,t)$, where $R_z(\Delta t)$ is the matrix representing constant-speed rotation about the $z$-axis after time $\Delta t$.
An observer in the IF who measures a particle at times $t$ and $\Delta t$ will see the particle move at some non-zero angle, $\eta$, to the local magnetic field.
The average speed measured in the IF is completely determined by the motion of the field line, which is known, and the angle $\eta$.
Indeed, the fact that the particle's speed in the IF is necessarily $v < c$ is equivalent to a finite range of allowed values of $\eta$, as illustrated in Fig. \ref{fig:beadonwire}.
Moreover, the extreme values of $\eta$ correspond to the limit $v \rightarrow c$.

To find which values of $\eta$ correspond to a given velocity, we take the vector sum illustrated in Fig \ref{fig:beadonwire} in the limit $\Delta t \rightarrow 0$.
In this limit, the vector $\overline{PP^\prime}$ becomes the particle's instantaneous velocity as measured in the IF, the vector $\overline{PQ}$ becomes the azimuthal velocity, and the vector $\overline{QP^\prime}$ is tangent to the local magnetic field line at $P$.
Thus, $\overline{PP^\prime} = \overline{PQ} + \overline{QP^\prime}$ becomes a vector triangle for which we know two sides (the particle's measured or assumed speed and the azimuthal velocity) and an angle (between the azimuthal direction and the local magnetic field), and which therefore can be solved completely.

As implied by Fig. \ref{fig:beadonwire}, there are in general two solutions for a given particle speed.
By solving the vector triangle for $v \approx c$, the normalised velocity vector can be expressed as
\begin{equation}
    \hat{\bm{v}} = \rho^\prime\hat{\bm{\phi}}
        + \left[
            -\rho^\prime(\hat{\bm{\phi}}\cdot\hat{\bm{B}}) \pm
            \sqrt{1 - (\rho^\prime)^2(1 - (\hat{\bm{\phi}}\cdot\hat{\bm{B}})^2)}
        \right] \hat{\bm{B}}
    \label{eqn:velocity}
\end{equation}
where $\rho^\prime = \rho/r_L$ is the azimuthal corotation speed normalised to the speed of light, $\rho$ is the point's perpendicular distance from the $\hat{\bm{z}}$-axis, $\hat{\bm{\phi}}$ is the azimuthal unit vector, and $\hat{\bm{B}}$ is the unit tangent to the local magnetic field in the inertial observer frame.
In most cases in the region of interest (i.e. above the polar cap), the positive solution corresponds to outward-flowing particles and the negative solution to inward-flowing particles.
Throughout the rest of this paper, we will consider only positive solution.
Moreover, because Eq. \eqref{eqn:velocity} is defined at all points in the magnetosphere, it describes a unique vector field (up to the sign of the radical), from which all other geometric properties can be directly derived, as is shown later in \S\ref{sec:beam_shape}.

The particle acceleration field can be directly computed from the velocity field.
Like the magnetic field, the velocity field ``corotates'' with the pulsar, and in the context of a time-dependent velocity field, the acceleration field must be identified with the material derivative
\begin{equation}
    \vec{\bm{a}} = \pd{\vec{\bm{v}}}{t} +
                   v_x\pd{\vec{\bm{v}}}{x} +
                   v_y\pd{\vec{\bm{v}}}{y} +
                   v_z\pd{\vec{\bm{v}}}{z}.
    \label{eqn:acceleration}
\end{equation}
Because the velocity $\vec{\bm{v}}$ has been constructed to have constant magnitude ($v \approx c$), the acceleration can be reduced to
\begin{equation}
    \vec{\bm{a}} = c\left(\pd{\hat{\bm{v}}}{t} +
                   v_x\pd{\hat{\bm{v}}}{x} +
                   v_y\pd{\hat{\bm{v}}}{y} +
                   v_z\pd{\hat{\bm{v}}}{z}\right).
    \label{eqn:acceleration1}
\end{equation}

Eqs. \eqref{eqn:velocity} to \eqref{eqn:acceleration1} were derived with an arbitrary magnetic field, but one must be careful to use only (time-dependent) magnetic fields that are physically meaningful in the IF.
For an investigation of first-order effects, a static rotating dipole is a sufficiently good approximation, but throughout this paper, and in the accompanying numerical code\footnote{PSRGEOM, obtainable from \url{https://github.com/robotopia/psrgeom}}, we have implemented the vacuum field derived originally by \citet{Deutsch1955} (cf \citealt{Arendt1998,Dyks2004a}).

\section{Construction of the beam}
\label{sec:beam_shape}

The salient feature of the velocity field as defined by Eq. \eqref{eqn:velocity} is that there is a unique emission direction associated with each emission location.
Therefore, assuming that magnetospheric propagation effects are negligible (which may very easily not be the case, e.g. see \citealt{Barnard1986a}), the beam shape can be predicted for a given set of emission regions.
A complete model also requires that the spectral output of each point is known, but in this analysis we will ignore any frequency dependence and only consider emission to be either on or off.

Before presenting the numerical beam shapes, we review the treatment of beam shapes in the slow rotation limit, and then derive the distortions introduced by finite rotation speeds.
In this and following sections, $(r,\theta,\phi)$ refer to spherical coordinates aligned with the magnetic axis.
That is, $r$, $\theta$, and $\phi$ are the radial distance, the magnetic colatitude, and the magnetic azimuth, respectively.

In the absence of rotation, the velocity field is everywhere parallel to the magnetic field, as can be confirmed by letting $r_L \rightarrow \infty$ (i.e. $\rho^\prime \rightarrow 0$) in Eq. \eqref{eqn:velocity}.
In this case, emission will be axisymmetric about the magnetic axis, $\hat{\bm{\mu}}$, and we can define the beam magnetic colatitude\footnote{Ordinarily, the term ``half opening angle'' is used for this quantity, which makes sense in the context of conal beam shapes, but here we use ``beam colatitude'' to emphasise the fact that the quantity is defined for all points in the magnetosphere, regardless of the global shape of the actual, observed beam.} to be $\Gamma = \cos^{-1}(\hat{\bm{B}}\cdot\hat{\bm{\mu}})$.
In the same limit, the Deutsch field reduces to a static dipole, and in this case, this colatitude becomes \citep[for a fuller treatment, see][]{Gangadhara2001,Gangadhara2004}
\begin{equation}
    \Gamma_\text{nr} = \tan^{-1}\left(\frac{3\sin\theta\cos\theta}{3\cos^2\theta-1}\right),
    \label{eqn:Gamma_nr}
\end{equation}
with the well known small angle approximation $\Gamma \approx (3/2)\theta$.
The subscript ``nr'' here indicates non-rotation.
The emission, like the magnetic field, will have no azimuthal component in the magnetic coordinate system.

With rotation, the important quantity for the beam shape is the angle between the velocity field and the magnetic axis, i.e. the deflected beam colatitude $\Gamma = \cos^{-1}(\hat{\bm{v}}\cdot\hat{\bm{\mu}})$.
The expression of this quantity in terms of spherical coordinates is a formidable exercise in algebra, but we can compare in more detail the more simplified case of an aligned rotator, $\alpha = 0$.
Then,
\begin{equation}
    \cos\Gamma = (3\cos^2\theta - 1)\sqrt{\frac{1 - (r^\prime\sin\theta)^2}{3\cos^2\theta+1}}.
    \label{eqn:aligned_Gamma}
\end{equation}
In the small angle approximation, we find
\begin{equation}
    \Gamma \approx \sqrt{(r^\prime)^2 + \frac94}\,\theta + O(\theta^3),
    \label{eqn:aligned_Gamma_approx}
\end{equation}
showing that the beam is dilated by an extra factor that depends on the emission height.

In the context of a given field line, $r^\prime$ and $\theta$ are coupled.
For a dipole geometry, the relationship between them is: $r \propto \sin^2\theta$, where the constant of proportionality is the field line's maximal distance from the origin.
Here, we follow the convention of \citet{Gangadhara2001} and others, and identify a field line by the magnetic colatitude of the point at which it pierces the pulsar surface, i.e. the ``footpoint'' colatitude $\theta_p$.
Moreover, we normalise the footpoint colatitude to the colatitude of the last open field lines, $\theta_L = \sin^{-1}\sqrt{r_p/r_L}$, so that the quantity $s \equiv \theta_p/\theta_L$ equals $0$ at the magnetic pole and unity at the polar cap radius.
By fixing $s$, we can make the substitution
\begin{equation}
    r^\prime \rightarrow \frac{r_p\sin^2\theta}{r_L\sin^2(s\theta_L)^2},
\end{equation}
in which case the approximation \eqref{eqn:aligned_Gamma_approx} reduces to $\Gamma \approx (3/2)\theta$ as before, with the $s$ term only making an appearance in the 5th order term in $\theta$.

As well as a slight expansion, the beam also acquires an azimuthal component, which due to axisymmetry also depends only on $r^\prime$ and $\theta$.
If the intensity of the pulsar beam is also axisymmetric, the azimuthal component becomes irrelevant (for the aligned rotator), since it only serves to rotate the beam around the common axis.
However, the intensity profile generated by a rotating carousel is (or may be) only axisymmetric after averaging over a carousel rotation.
In order to appreciate the effects of aberration on the global beam shape, we momentarily neglect the effects of carousel rotation by considering the slow rotation limit (i.e. $P_4 \rightarrow \infty$).
In this limit, each beamlet originates from a narrow tube of field lines emerging from a discrete spark at the tube's base.
The beam shape due to a single narrow tube will in general be some form of spiral, as emission from different heights along the tube is deflected by different amounts.
Of course, the whole spiral will not necessarily be present because the emission is presumably only generated within a finite range of heights at a given frequency.
Nevertheless, for a given line of sight cut, it may be the case that a large enough range of heights are sampled that the map produced by the cartographic transform reveals the spiral nature of the individual beamlets.

\begin{figure}
    \centering
    \includegraphics[scale=0.3]{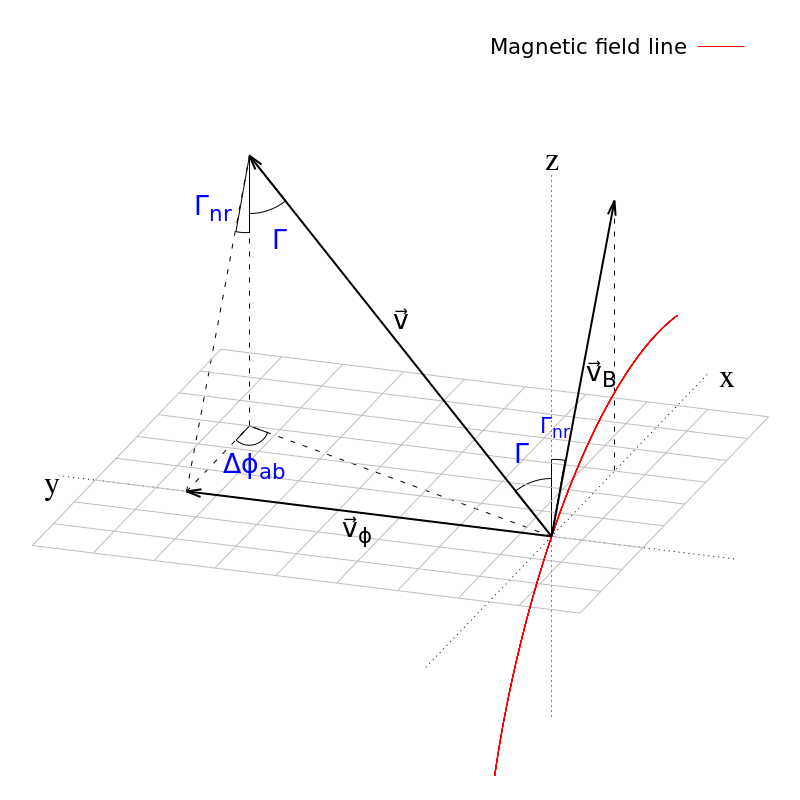}
    \caption{The geometry of rotationally deflected emission for an aligned rotator. The red line is a magnetic field line in the $xz$-plane. The vector $\vec{\bm{v}}$ is deflected from the tangent to the magnetic field line, such that $\Gamma > \Gamma_\text{nr}$ and $\Delta\phi > 0$. The vectors $\vec{\bm{v}}_\phi$ and $\vec{\bm{v}}_B$ correspond to the first and second terms of Eq. \eqref{eqn:velocity}, respectively.}
    \label{fig:alignedcase}
\end{figure}

To find the beam's azimuthal component (in the aligned case), we first compute the quantity $\Delta\phi_\text{ab} = \tan^{-1}(-v_y/v_x)$.
Axisymmetry allows us to calculate the azimuthal aberration by simply evaluating $\Delta\phi_\text{ab}$ in the $xz$-plane (i.e. when $\phi = 0$).
This gives
\begin{equation}
    \tan(\Delta\phi_\text{ab}) = -\frac{r^\prime}{3\cos\theta}\sqrt{\frac{3\cos^2\theta + 1}{1 - (r^\prime\sin\theta)^2}}.
    \label{eqn:aberration}
\end{equation}
The above result may be more simply derived by noting that in Fig. \eqref{fig:alignedcase}, the common sides of the triangles formed by $\vec{\bm{v}}$, $\vec{\bm{v}}_\phi$, and $\vec{\bm{v}}_B$ must be related by
\begin{equation}
    \cos(\Delta\phi_\text{ab}) = \frac{\tan\Gamma_\text{nr}}{\tan\Gamma},
\end{equation}
and then substituting in the Eqs. \eqref{eqn:Gamma_nr} and \eqref{eqn:aligned_Gamma}.

Having obtained a complete description of the deflected emission beam, we must now incorporate the effects of retardation into the analysis.
Similarly to $\Gamma$ above, we can define the retardation angle, $\Delta\phi_\text{ret}$, as a valid quantity at every point in the magnetosphere, regardless of whether or not the observer is in the direction to observe it:
\begin{equation}
    \Delta\phi_\text{ret} = -r^\prime (\hat{\bm{r}}\cdot\hat{\bm{v}}),
    \label{eqn:retardation}
\end{equation}
which, for an aligned rotator becomes
\begin{equation}
    \begin{aligned}
        \Delta\phi_\text{ret}
            &= -2r^\prime\cos\theta\sqrt{\frac{1 - (r^\prime\sin\theta)^2}{3\cos^2\theta+1}} \\
            &= -\frac{2r^\prime\cos\theta\cos\Gamma}{3\cos^2\theta - 1}
    \end{aligned}
\end{equation}
The negative sign serves the same purpose as that in Eq. \eqref{eqn:aberration}, namely, to shift the observed emission to earlier rotation phases.
This shift is equivalent to a further azimuthal distortion of the overall beam pattern, which in the aligned case serves to enhance the curvature of the spiral shape of the beam.
For a dipolar field, Eq. \eqref{eqn:retardation} reduces to the approximation $\Delta\phi_\text{ret} \approx -r^\prime + O({r^\prime}^3)$, in agreement with \citet{Dyks2004}.

\begin{figure}
    \centering
    \includegraphics[scale=0.4]{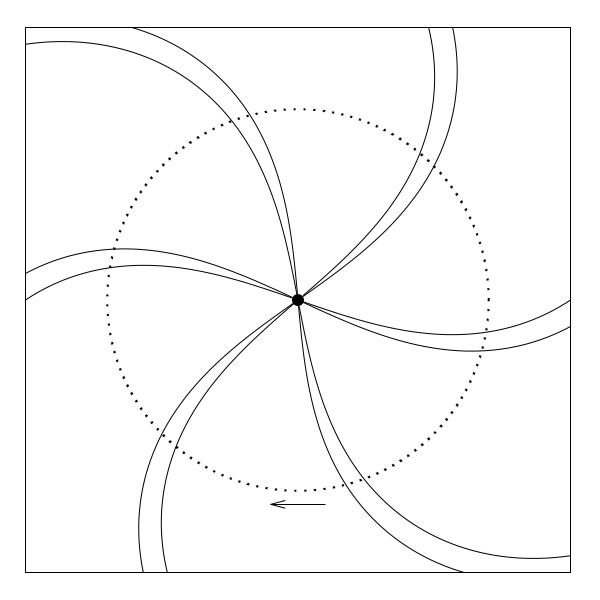}
    \caption{A schematic diagram of the spiral structure of beamlets arising from carousel sparks for an aligned rotator---once rotation effects and retardation have been factored in---if the emission were allowed to come from all heights. Note that the beam pattern itself would only be axisymmetric for a perfectly aligned rotator. In general, the line of sight (dotted line) cuts through the beamlets at an oblique angle that depends on the emission height, visible as an azimuthal distortion of the beam pattern. The line of sight cuts from right to left, as indicated by the arrow. The directional sense of the distortion is the same as the sense of stellar rotation, i.e. clockwise when viewed from above.}
    \label{fig:spiral}
\end{figure}

By choosing a fixed $s$, the equations above constitute a complete description of the spiral beam pattern, as illustrated in Fig. \ref{fig:spiral}.
The foregoing construction is also valid (albeit algebraically daunting) for pulsars of arbitrary inclination angle $\alpha$.

Emission from field lines with smaller $s$ suffer from greater azimuthal distortion.
This makes intuitive sense: the curvature of these field lines is smaller, and so the particles must climb to greater heights in order to produce beams of equivalent apparent magnetic colatitude.
In the case of any one particular pulsar, however, it generally remains unknown exactly which magnetic field lines are the active ones.
On the other hand, if azimuthal distortion of the beam can be measured, and assuming there are no significant additional distorting effects, the footpoints colatitudes can be constrained.

\section{Simulating pulsestacks}
\label{sec:pulsestacks}

A true comparison of the predictions of the present model and the polar maps generated by the cartographic transform requires the simulation of pulsestacks which can then be transformed.
In this section, we describe in detail the process of producing simulated pulsestacks from the geometric model outlined above.

Clearly the geometric framework alone is insufficient to simulate observed pulsestacks; it can only say when emission from a given location in the magnetosphere would be seen by the observer, not how bright that emission would be at a given frequency.
However, the primary consideration here is the morphology of the drift bands, i.e. when the subpulses in successive pulses arrive in relation to each other.
By virtue of the underlying assumption of the carousel model---that the emissivity of a field line is directly related to the spark activity at its base---we can use a simple model for estimating the observed intensities from specific emission locations in which they are proportional to the spark activity at their respective footpoints.

For an observer whose line of sight makes an angle $\zeta$ to the rotation axis, the only points on a given magnetic field line that are potentially visible are those for which
\begin{equation}
    \hat{\bm{v}}\cdot\hat{\bm{\Omega}} = \cos\zeta
    \label{eqn:visiblepoint}
\end{equation}
(in the aligned case, this condition is equivalent to $\Gamma = \zeta$).
When $\zeta > \alpha$, there is a unique height on each magnetic field line above the polar cap where this condition is satisfied.
On the other hand, when $\zeta < \alpha$, field lines on the equatorial side of the magnetic axis start at the surface with associated velocities that make an angle to the rotation axis $> \alpha$, and this angle only increases at greater emission heights as the field line drops away, never becoming equal to $\zeta$.
Then $\zeta < \alpha < \cos^{-1}(\hat{\bm{v}}\cdot\hat{\bm{\Omega}})$, so Eq. \eqref{eqn:visiblepoint} can never be satisfied and these field lines never come into view of the observer.
Field lines on the poleward side, however, initially curve ``upwards'' and then ``backwards'', so that the angle $\cos^{-1}(\hat{\bm{v}}\cdot\hat{\bm{\Omega}})$ starts at some value $< \alpha$, decreases through $\zeta$ until some minimum value is reached, and then increases again as the magnetic field drops away on the other side.
These field lines therefore include visible points at two different heights, whose exact values can be found by solving Eq. \eqref{eqn:visiblepoint} numerically.
Since the second visible point is typically much higher than the first and appears near the anti-fiducial point, they are not expected to be observed in type $S_d$ pulsars (i.e. whose profiles consist of a single conal component) and we hereafter consider only the lower visible point.

The fact that each visible magnetic field line is only observed at most at one height implies that \emph{any} arbitrary intensity profile or pulsestack can be obtained by a careful choice of (dynamic) spark pattern on the surface.
However, we note that multiple field lines can contribute to the emission observed at a given phase, so that at all phases a range of heights is (potentially) observed.
In this geometric model, having a range of emission heights is necessary to produce a spread in the polarisation angle at a given rotational phase, because just as each point in the magnetosphere has associated with it a unique velocity from Eq. \eqref{eqn:velocity}, each point also has a unique acceleration vector from Eq. \eqref{eqn:acceleration}, and therefore a unique polarisation angle.
Thus, a realistic model requires that a full, two dimensional spark pattern is used, in order that a range of heights and therefore a range of polarisation angles are sampled at each phase.
However, for ease of computation, the simulated pulsestacks presented here are generated from a one-dimensional carousel at a fixed value of $s$ whose ``sparks'' have Gaussian profiles in magnetic azimuth.
By choosing only a single $s$ value and making the rest of the polar cap ``inactive'', we are forcing the polarisation angle to adopt a unique value at each phase---that is, every pulse will have the exact same sweep of polarisation angles---which we will interpret as the ``average'' polarisation angle.

The steps to create a pulsestack are then as follows.
For each footpoint in the (one dimensional) carousel, the field line is traversed numerically using Runge-Kutta integration until a point is found that has the property that $\hat{\bm{v}}\cdot\hat{\bm{\Omega}} = \cos\zeta$, where $\hat{\bm{v}}$ is calculated from Eq. \eqref{eqn:velocity}.
This is the so-called ``visible point'' of that field line.
The polarisation angle associated with the visible point is then calculated by projecting the acceleration vector of Eq. \eqref{eqn:acceleration} onto the observer's sky plane (i.e. normal to $\hat{\bm{v}}$).
This is done to verify that the chosen parameters are consistent with the observed (average) polarisation swing.

The observed phase is determined by calculating $\Delta\phi_\text{ab} + \Delta\phi_\text{ret}$.
The first term gives the rotation phase at which the emission must have taken place, and the second term is the retardation shift due to photon flight times.
Next, we find the rotation phase at which a spark event at the field line's footpoint would have occurred in order that the resulting particle stream (assumed to be travelling at ultra-relativistic speeds) would reach the visible point at phase $\Delta\phi_\text{ab}$,
\begin{equation}
    \Delta\phi_\text{sp} = \Delta\phi_\text{ab} - \ell^\prime,
    \label{eqn:sparkphase}
\end{equation}
where $\ell^\prime \equiv \ell/r_L$, and $\ell$ is the distance traversed by a particle as it corotates with the pulsar and climbs from the surface to the emission point, which is calculated numerically at the same time that the visible point is being found.
For pulse number $p$ in the pulsestack, $\Delta\phi_\text{sp}$ is converted (via the pulsar rotation period, $P_1$) into a time $t_\text{sp} = P_1(p + \Delta\phi_\text{sp}/2\pi)$, which can then in turn be converted into an intensity via the carousel model
\begin{equation}
    I(t_\text{sp},\phi) \propto \sum_{n=1}^N \exp\left[-\frac{(\phi - 2\pi(\frac{n}{N} + \frac{t_\text{sp}}{P_4}))^2}{2\sigma^2}\right],
\end{equation}
where $\phi$ is the magnetic azimuth of the footpoint and $\sigma$ parametrises the angular width of the individual spark profiles.

\subsection{Phase-dependent intensity modulation}

\begin{figure}
    \centering
    \includegraphics[scale=0.3]{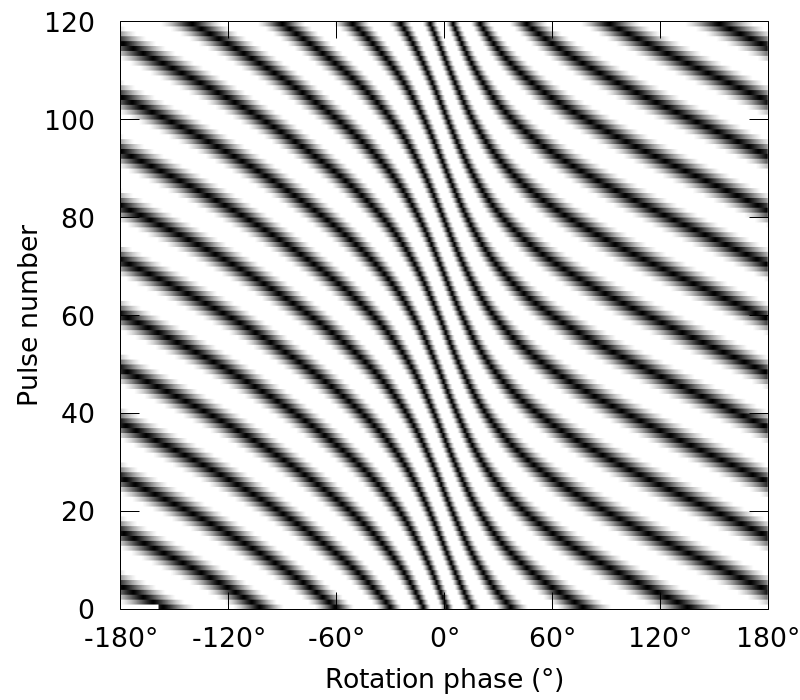}
    \caption{A simulated pulsestack showing drift bands across all $360^\circ$ of rotation phase, using parameters drawn from one of the proposed viewing and carousel geometries of PSR B0809+74 (see \S\ref{sec:beamlets} for details of the chosen parameters).}
    \label{fig:pulsestack360}
\end{figure}

The foregoing procedure will produce a pulsestack where emission is ``observed'' across all $360^\circ$ of longitude, regardless of the local conditions at the emission point, as illustrated in Fig. \ref{fig:pulsestack360}.
In other words, it follows the assumption of the carousel model that spark activity at the footpoint of a field line is a necessary, but not a sufficient condition for coherence to occur on that field line.
Some extra condition, presumably connected to the emission mechanism, must therefore be imposed in order to decide how the drift bands are modulated over pulse phase.

If a mechanism such as particle bunching is assumed (as first suggested by \citealt{Ruderman1975}, but see, e.g., \citealt{Melrose2017} for arguments against bunching as a viable mechanism), then the beam intensity is proportional to $N^2$, where $N$ is the number of particles in a ``bunch'' (i.e. within a volume element with its linear dimension smaller than the emitted wavelength in the co-moving frame).
In this case, the observed emission pattern for curvature radiation from a single particle can be calculated, and scaled up by the square of the local plasma density at each emission site.
\citet{Thomas2010} describe one possible method for doing such a calculation, although it should be noted that they incorporate the phase-shifting effects of aberration and retardation post hoc.
To our knowledge, no similar method exists for estimating intensity profiles assuming other possible mechanisms, such as maser mechanisms \citep{Ginzburg1975} or plasma wave instabilities \citep{Cheng1977}.

It should be realised, however, that the positions of the drift bands in observed rotation phase are, under the assumptions of the carousel model, agnostic to which emission mechanism is ultimately adopted\cmmnt{\footnote{This statement does not allow for the minor apparent shift caused by the profile modulation itself, which will serve to skew the drift bands slightly towards the centres of the profile components.}}.
If, therefore, the goal is to study how the drift bands are affected by the geometry of the magnetosphere, one can avoid (or at least defer) the unenviable task of choosing and modelling one particular coherent emission mechanism at the expense of the others.
If a comparison with real data is still needed, it suffices to modulate the simulated pulsestack with an actual average profile at the desired frequency, which would include asymmetries, if any, resulting from both the emission physics and geometrical considerations.

\section{Observable predictions of the geometric model}
\label{sec:applications}

\subsection{Drift band morphology}

The drift bands in Fig. \ref{fig:pulsestack360} show clear and significant curvature over the full range of rotation phases, due to the phase-dependent rate at which the visible point cuts across the carousel beam at different rotation phases.
From the construction outlined in the previous section, the drift bands as they appear on the pulsestack have the dual properties \citep[common to pulsars with drifting subpulses, as stated in][and references therein]{Edwards2002} that the horizontal ``distance'' between them (i.e. the time between successive subpulses, $P_2$) is a pure function of pulse phase, and the vertical ``distance'' ($P_3$) is a pure function of pulse number.
If aliasing is present, the vertical spacing of the drift bands will be $P_3 \ne \hat{P}_3 = P_4/N$ (see, e.g., \citealt{Deshpande2001} for a detailed discussion), but this will not affect the constancy of either $P_3$ or $\hat{P}_3$ as long as $P_4$ and $N$ remain constant (which is not always the case, e.g., \citealt{McSweeney2017}).

The shape of each simulated drift band is identical.
This, of course, precludes the possibility of accurately modelling pulsars whose subpulse drifting properties undergo any kind of evolution over time, such as nulling, mode switching, or a changing drift rate \citep[see][for an example of all three occurring in the same pulsar, PSR B0031-07]{McSweeney2017}.
In these cases, it is assumed that $N$, $P_4$, $s$, or some combination of them are not constant in time \citep[e.g.][]{Smits2005}.

\subsection{Appearance of reconstructed beamlets}
\label{sec:beamlets}

The simulated pulsestacks can be subjected to all the standard analyses used to study subpulse drifting.
In particular, they are amenable to the reconstruction of polar beam maps via the cartographic transform of \citet{Deshpande2001}.
As a case study, we investigate one of the viewing geometries proposed by \citet{Rankin2006} for PSR B0809+74, whose transformed beam map shows a remarkable azimuthal skewing of the beam maps (see their Fig. 4).
The viewing geometry assumed $\alpha = 9^\circ$, $\zeta = 13.5^\circ$, and number of sparks $N = 10$.
Even though the authors emphasise the difficulty of distinguishing between many sets of viable parameters, they argue that in this case aliasing is not present, in which case $P_3 \approx 11.1\,P_1$ is the time interval between successive drift bands at a fixed rotation phase, giving $P_4 = N\times P_3 \approx 143\,$s.

\begin{figure}
    \centering
    \includegraphics[scale=0.25]{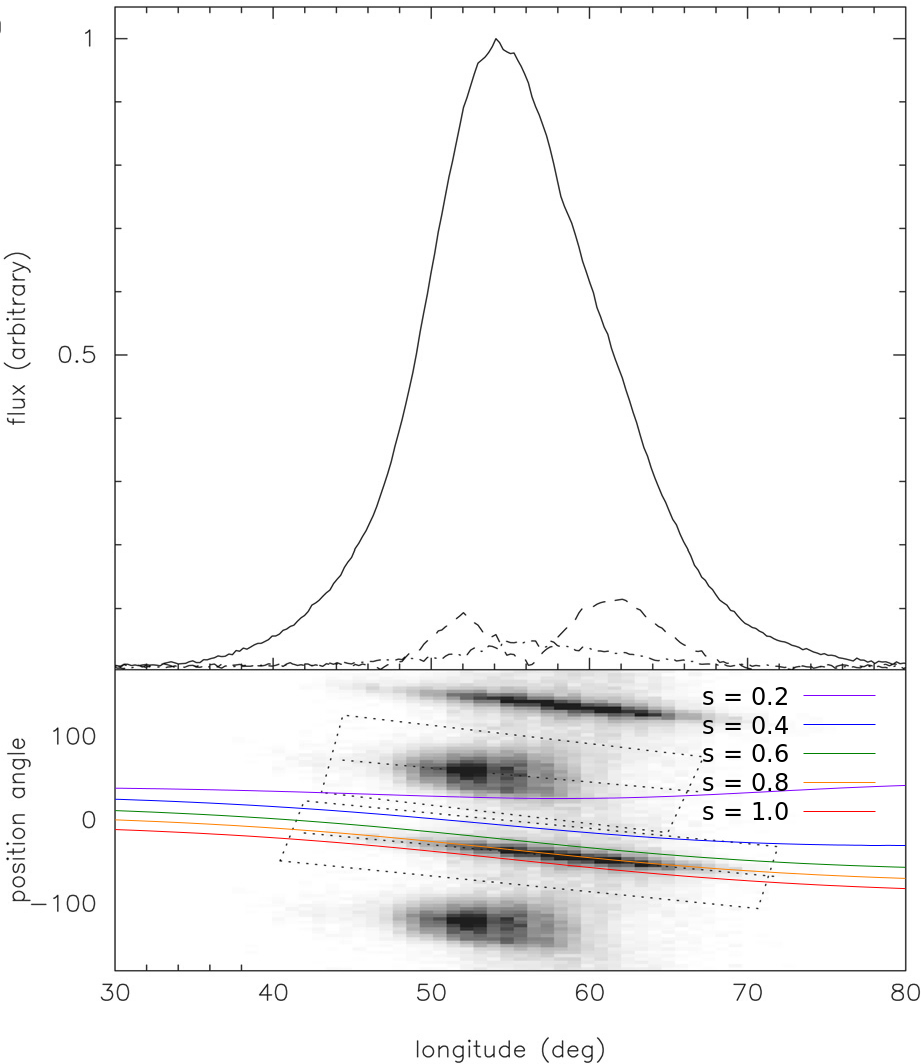}
    \caption{The pulse profile and polarisation angle histogram of B0809+74 at $328\,$MHz, reproduced from \citet{Rankin2006a}, and overlaid with the predictions of the polarisation angle curves for various values of the normalised radius of the carousel, $s$. The vertical offsets are arbitrary.}
    \label{fig:pa}
\end{figure}

One parameter remains free: the (normalised) radius of the carousel, $s$.
We first tested which values in the range $0 < s < 1$ produced polarisation angle curves consistent with the observations (for one of the orthogonal polarisation modes only).
Fig. \ref{fig:pa} shows the polarisation angle curves predicted by the geometric model.
As expected, values for relatively large $s$ agree better, as these correspond to relatively low emission heights, at which the geometric model more closely approximates the rotating vector model, which was employed in the determination of the viewing geometry parameters.
In particular, we note that values in the range $0.4 \lesssim s \lesssim 1.0$ agree sufficiently well with the polarisation angles in the pulse window.

\begin{figure}
    \centering
    \includegraphics[scale=0.30]{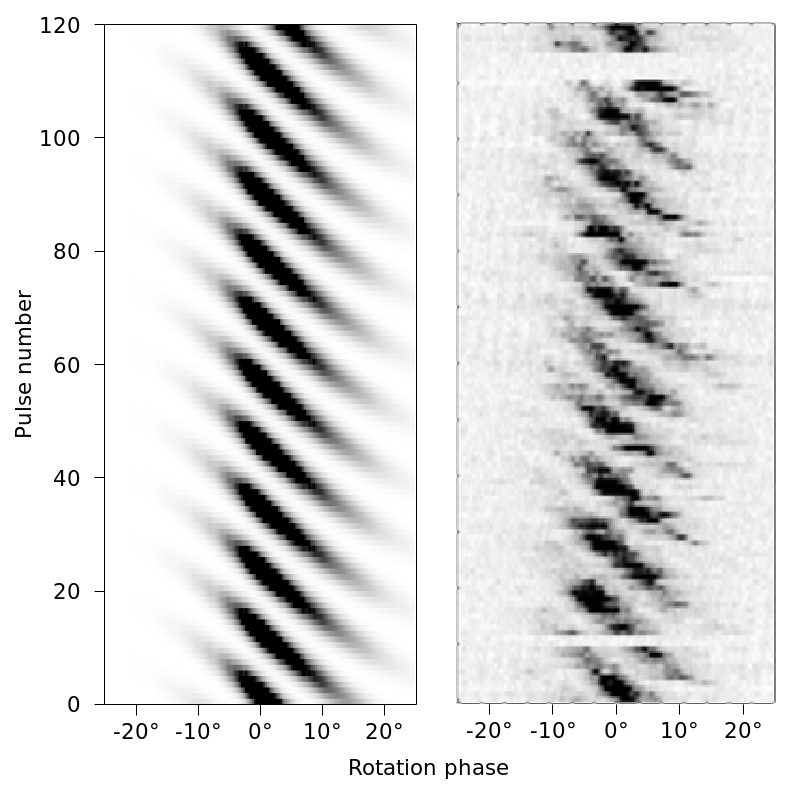}
    \includegraphics[scale=0.30]{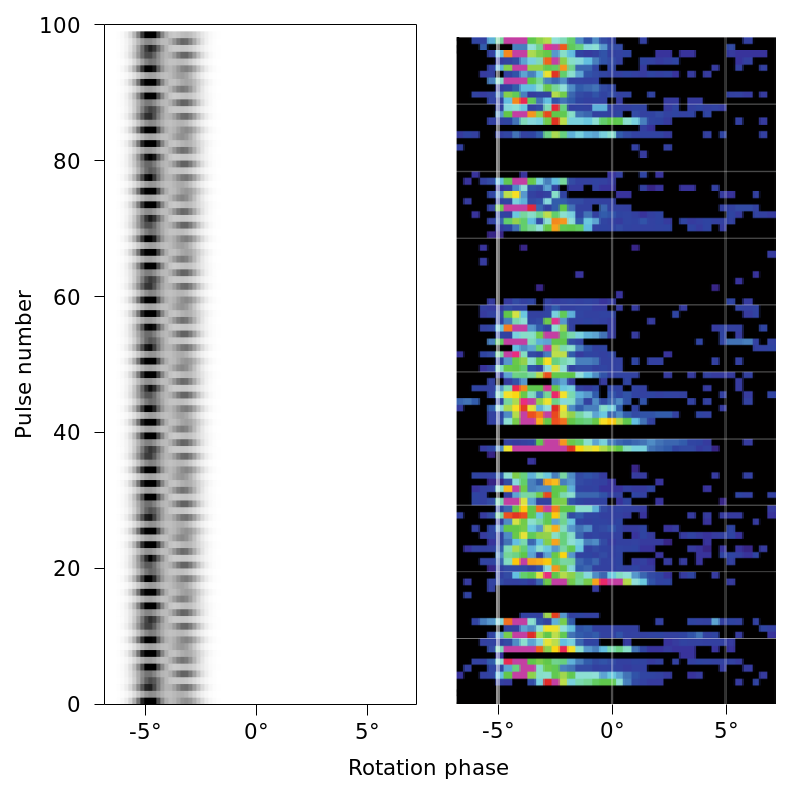}
    \caption{(Top left) A simulation of PSR B0809+74's drift bands, generated using the parameters given in \citet{Rankin2006}, assuming $s = 0.5$ and spark size $\sigma = 6^\circ$. The pulsestack was modulated with a $150\,$MHz profile from the European Pulsar Network database (see text for details). The drift bands are distinctly (albeit slightly) curved, in agreement with observation. (Top right) A comparable pulse sequence from an observation of B0809+74 taken at $313\,$MHz \citep{Gajjar2014}. The sequence contains several short null sequences, which the simulation lacks. (Bottom) Same as above, but for pulsar B2034+19 (see text for chosen parameters). The modulating profiles for the two components (at $-5^\circ$ and $-3.5^\circ$) were constructed from the profile given in \citet{Rankin2017a}.}
    \label{fig:simulated-pulsestacks}
\end{figure}

After choosing a value for $s$, the pulsestack can be simulated according to the procedure described in Section \S\ref{sec:pulsestacks}.
Fig. \ref{fig:simulated-pulsestacks} shows a simulated intensity pulsestack for pulsars B0809+74 and B2034+19, using parameters reported in \citet{Rankin2006} and \citet{Rankin2017a} respectively.
The average profile used to modulate the pulsestack was a $\sim150\,$MHz profile retrieved from the European Pulsar Network database\footnote{\url{http://www.epta.eu.org/epndb/}}, supplied by \citet{Noutsos2015}.
The authors assigned the fiducial point in their profile by means of a Rotating Vector Model \citep[RVM;][]{Radhakrishnan1969,Komesaroff1970} fit to the observed polarisation curve.
Since (by virtue of Fig. \ref{fig:pa}) the simulated polarisation curve is sufficiently close to the RVM for the range $0.4 < s < 1.0$, the association with their empirically determined fiducial point with the simulated fiducial point is justified.
Varying $s$ within this range does not significantly alter the shape of the drift bands.

\begin{figure*}
    \centering
    \includegraphics[scale=0.4]{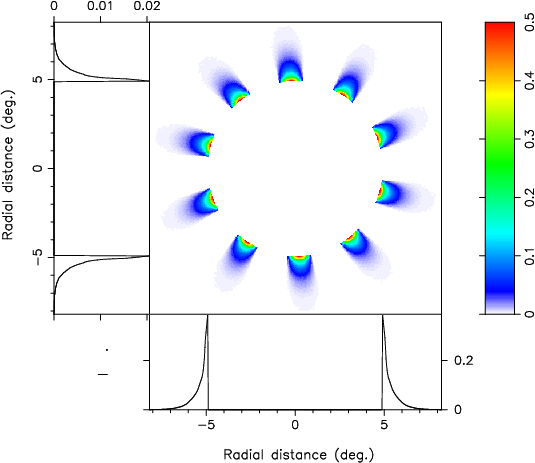}
    \includegraphics[scale=0.4]{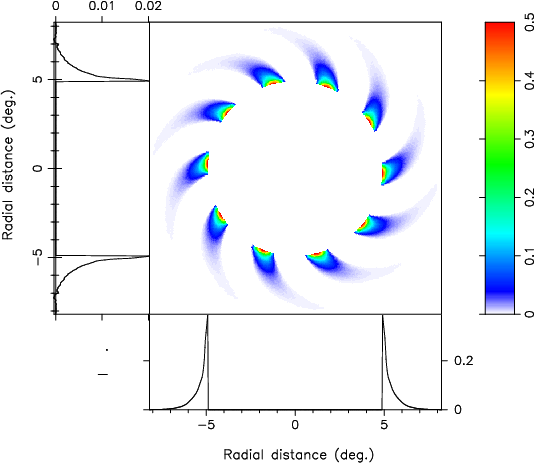}
    \caption{Two polar maps made from simulated data. On the left, $s=0.5$ was used, and on the right, $s=0.25$. The azimuthal distortions, argued here to be primarily due to the greater aberration associated with greater emission heights, are clearly present in the right-hand plot, but \emph{have the opposite sense to that seen by \citet{Rankin2006a}} (cf. their Fig. 4).}
    \label{fig:polarmap}
\end{figure*}

Having obtained a simulated pulsestack, we then applied the cartographic transform in order to obtain a simulated polar map of the pulsar beam.
Fig. \ref{fig:polarmap} shows the results for two choices of footprint polar radius, $s=0.5$ and $s=0.25$.
When $s$ values are chosen in the range $0.4 \le s \le 1.0$, no significant azimuthal distortions were observed, but smaller values of $s$ yield very noticeable azimuthal distortions with similar characteristics as that observed by \citet{Rankin2006}.
Curiously, the direction of the azimuthal distortions differ, with the simulated beamlets being skewed in the counter-clockwise (CCW) direction, while the real data is skewed in the clockwise (CW) direction.
This is discussed further in the next section.

\section{Discussion}
\label{sec:discussion}

\subsection{Comparison of simulated and real data}

The geometric model described in this work is capable of producing simulated pulsestacks from a set of parameters describing the viewing geometry and the underlying carousel of sparks near the surface.
We have assumed that the appearance and morphology of drift bands is entirely determined by which field lines are ``active'' (i.e. those whose footpoints coincide with a spark event at a given time), and that the coherent emission mechanism itself is only responsible for how the whole drift band pattern is modulated over $360^\circ$ of rotation phase.
The unknown details of the emission mechanism have not been investigated here; rather, we have modulated our simulated pulsestacks with the profiles of real data taken at some desired frequency.

Our purpose has been to show that the geometric model is sufficient to reproduce the essential characteristics of the drift band patterns actually observed, without committing to any one particular coherent emission mechanism.
The two simulated pulsestacks in Fig. \ref{fig:simulated-pulsestacks} illustrate the application of the geometric model to two pulsars with very different viewing geometries.
Except for the absence of nulls, the simulated pulsestacks exhibit very similar morphologies to their real counterparts; in particular, the similar curvature of the drift bands of B0809+74 and the odd-even subpulse modulation over the pulse sequence in B2034+19 \citep[discussed in][]{Rankin2017a}.

We have not attempted an exhaustive search in these case studies for the best fitting carousel parameters.
For the latter pulsar, we have adopted the following parameters: $N=5$, $P_4=6.65\,$s, spark size $\sigma = 20^\circ$, and that the direction of the carousel motion has the opposite sense to the rotation of the pulsar.
{Assuming the double cone model of \citet{Rankin2017a}, we modelled the two components (at approximate rotation phases $-5^\circ$ and $-3.5^\circ$) separately, assigning them carousel radii of $s=1$ and $s=0.8$ respectively.
In order to approximate the profile contributions from the relevant drift mode, we fit four Gaussians to the total profile and discarded the two rightmost components, which appear to come mainly from the other drift mode.
The sparks of the inner cone were set midway in magnetic azimuth between the sparks of the outer cone.
Other sets of parameters were found to result in a similar morphology, and a comprehensive search for the best fitting set should include, for example, a comparison of the fluctuation spectra.
This, however, is beyond the scope of the present work, primarily aimed to highlight the wide variety of simulated pulsestacks that are possible with the geometric model.
A comprehensive model of B2034+19's drift bands in both its modes (here, only the first mode was simulated) is deferred to a future work.

\subsection{Effect on the interval between successive subpulses ($P_2$)}

An important feature of the geometric model is the inclusion of the time taken for the spark information to reach the visible point, a non-negligible time, which to our knowledge is entirely lacking in other models of carousel behaviour.
The quantity $P_2$, which measures the elapsed time between the observation of two adjacent sparks, must therefore include not only AR effects and continuous carousel rotation (affecting the \emph{apparent} spacing between the sparks, a theme discussed in \citealt{Yuen2016}), but also the difference of path lengths of the respective particle trajectories.
The length of the trajectories traced by particles in the IF ($\ell^\prime$) are necessarily different from the path length along the magnetic field lines on which they reside---rotation effects should not be neglected.

As described above, the trajectory lengths are calculated numerically, but we can get a sense of the magnitude and behaviour of the effect by taking the first-order approximation $\ell^\prime \approx r^\prime$.
Then the interval between the arrival of successive subpulses will depend on how far the carousel rotates in time $\sim \Delta r/c$, which in radians is $\frac{\Delta r^\prime P_1}{P_4}$.
On the leading side of the fiducial point, the emission heights start large, approach a minimum near the fiducial point, and grow large again on the trailing side (see Fig. \ref{fig:emissionheights}).
The inclusion of trajectory lengths therefore either serves to artificially augment or diminish the observed value of $P_2$ during leading phases (depending on whether the line of sight cuts across the beam pattern in the same or opposite directional sense, respectively, as the carousel rotation) and to have the opposite effect on $P_2$ on the trailing side.
This is similar to retardation, except that retardation always compresses the observed emission on the trailing side \citep{Dyks2010}.
Therefore, the present effect can either enhance or suppress the effect of retardation, depending on the relative signs of $P_1$, $P_4$, and $\beta = \zeta - \alpha$.

\subsection{Effect of finite spark size}

A similar comparison can be made between sparks with the same magnetic azimuth but with different magnetic colatitudes, i.e. different values of $s$.
In this case, we can assume to first order that the aberration angles are the same ($\Delta\phi_{\text{ab},1} \approx \Delta\phi_{\text{ab},2}$), and that therefore any difference in arrival time is due to the combined difference of particle trajectory length and photon flight time (retardation).
Here, second order effects become necessarily important, because the inclusion of only first order effects implies that the total path length of the spark information is the sum of the distance from the surface to the emission point, $\sim r$, and the distance from the emission point to the observer, $\sim(D-r)$ (where $D$ is the distance between the centre of the pulsar and the observer), which reduces to the constant distance $D$.
In reality, if the sparks span a large enough range of $s$ values, the inner parts of the spark (i.e. nearest the magnetic pole) may be ultimately observed measurably later in phase than the outer parts, which would be observed as a broadening of the drift bands in phase.

It has been suggested that changing drift modes is caused by a change in the radius of the spark carousel of PSR B0031-07 \citep{Smits2005}.
In their model, the observed difference of emission heights between the two most prominent drift modes is a function of observing frequency.
If this is the case, then the geometric model predicts that there should be a different rotational phase shift between the drift bands of each mode viewed simultaneously across a sufficiently wide frequency band, which has indeed been observed \citep{McSweeney2017a}.

Similar frequency-dependent effects have also been observed in PSR B0809+74 by \citet{Hassall2013} and PSR B0943+10 by \citet{Bilous2018}.
The latter argue that the evolution (in phase) of the drift bands can be understood in terms of the radius-to-frequency mapping (RFM; whereby the opening angle of the beam emerging from the same surface sparks increases at lower frequencies), affecting the projected phases at which the beamlets pass through the line of sight.
Our model is in principle consistent with RFM, but requires the implementation of two-dimensional surface sparks in order to make a qualitative comparison with the work of \citet{Bilous2018}, since in our (one-dimensional) model each field line is associated with only a single visible point at a geometrically determined height.

For the same reason, our model also predicts that (broadband) microstructure would \emph{not} show a frequency-dependent phase separation, assuming that the emission regions associated with micro-pulses span a sufficiently small number of field lines---again, because the location of the visible point (and thereby the degree to which AR effects are present) of a given emission column is independent of observing frequency.
Thus, the geometric model predicts that subpulses appear to move about in phase when viewed simultaneously across a wide frequency range (because different field lines are sampled), whereas the individual micro spark events that make up a spark ``patch'', if they do indeed occur on a small fixed set of field lines, would appear at the same phase at all frequencies.
Since B0031-07 is known occasionally to emit particularly bright pulses \citep{Tuoheti2011,Nizamdin2011}, it may provide a way to test the present geometric model, if sufficiently high time resolution observations of bright microstructure across a wide frequency band can be obtained \citep[see, e.g., the temporal stability of Crab microbursts across $\sim 2$ octaves of frequency, presented in Fig. 2 of][]{Hankins2016}.

\subsection{The rotational asymmetry of B0809+74's reconstructed beam}

We have shown that it is possible to simulate pulsestacks that, when the cartographic transform is applied to them, appear azimuthally skewed in a way that is at least qualitatively expected from AR effects.
Although this exercise was motivated by the azimuthal asymmetries observed by \citet{Rankin2006} for B0809+74, there are two major reasons why the effect described in this paper cannot provide an adequate explanation for their particular case.

The first reason is the direction of the asymmetry: \citeauthor{Rankin2006}'s beamlets appear to be skewed in the CW direction, whereas the simulated polar maps are skewed CCW.
Both polar maps were produced with the same set of parameters, which assume that the line of sight is cutting through the beam from right to left, with $\zeta > \alpha$.
In this scenario, height-dependent aberration always tends to shift the observed emission towards earlier rotation phases, which equates to the beamlets being skewed towards the right at the bottom of the polar map, with the net result of a CCW skew.

The second reason is the failure to find a consistent value of $s$ that can produce both the correct polarisation angle curve ($s \gtrsim 0.4$) and the azimuthal asymmetry ($s \lesssim 0.3$).
For the geometry of B0809+74, the former condition is equivalent to the restriction $r^\prime \lesssim 0.02$ ($r \lesssim 100\,$km), and the latter, $r^\prime \gtrsim 0.04$ ($r \gtrsim 250\,$km), which is demonstrated by the solid height curves in Fig. \ref{fig:emissionheights}.
\begin{figure}
    \centering
    \includegraphics[scale=0.3]{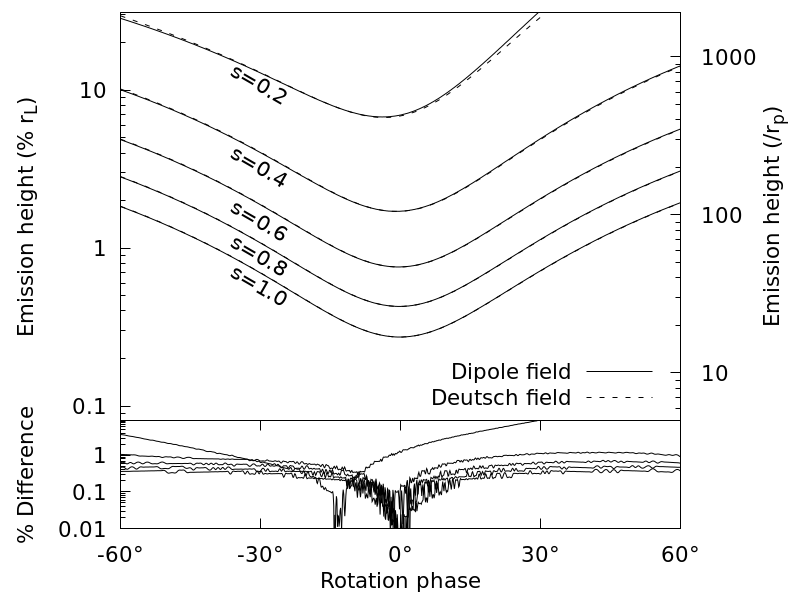}
    \caption{\emph{Top}: Numerical emission heights for B0809+74's proposed geometry, for both a dipolar magnetic field and a Deutsch field, assuming a stellar radius of $10\,$km. The left and right $y$-axes are equivalent, but presented in terms of the light cylinder radius, $r_L$, and the pulsar radius, $r_p$, respectively. \emph{Bottom}: The absolute percentage difference between the dipolar and Deutsch derived heights, normalised to the dipolar height. At the lower heights ($s \ge 0.4$), the two are almost indistinguishable ($< 1\%$ difference). The two coincide at a height-dependent phase, with the $s = 0.2$ curve showing the left-most minimum. The roughness of the curves is due to numerical noise in the PSRGEOM algorithm.}
    \label{fig:emissionheights}
\end{figure}
This difficulty is not necessarily insurmountable, because the RVM implicitly assumes sufficiently low heights that rotational effects are negligible.
As can be seen in Fig. \ref{fig:pa}, the polarisation angle curve tends to flatten when lower values of $s$ (and therefore greater emission heights) are chosen.
Therefore, there may well be a viewing geometry for the which the RVM would predict a steeper polarisation angle curve, but which agrees with the data for a sufficiently low value of $s$ that would make the azimuthal asymmetry significant.
If a change in the viewing geometry is required, then there is no reason to continue assuming that the other parameters of \citeauthor{Rankin2006}'s model are correct, including the number of sparks in the carousel, and the degree of drift band aliasing (currently assumed to be zero).
Finding such a geometry is beyond the scope of the present work, but could, in principle, be found by extending the RVM to include the rotational effects implicit in the equation for the acceleration field, Eq. \eqref{eqn:acceleration}.

One major difference between the analysis of \citet{Rankin2006} and the present work is that the azimuthal asymmetry was seen in only one OPM, whereas our analysis assumes (incorrectly) that all the observed emission belongs to the same OPM.
The polarisation plots in Fig. 4 of \citet{Ramachandran2002} suggest that the dominance of the OPMs depends very strongly on where in the drift band the emission in question occurs.
Choosing only one OPM is therefore tantamount to significantly changing the appearance of the drift bands, which will inevitably affect the appearance of the transformed beamlets.
In B0809+74's case, the way the OPMs are divided in the pulsestack would make the apparent drift rate larger (i.e. the drift bands more horizontal).
However, experimenting with the simulation parameters shows that one of the effects of increasing $s$ (but leaving the other model parameters the same) is to increase the apparent drift rate in the pulsestack, which would lead us to expect that selecting just one OPM would cause the polar map to increase in skewness in the CCW direction, opposite to what is observed.
Furthermore, the other OPM's polar map skewness would be similarly boosted because its drift bands would also be flattened---this is also not observed.
In summary, we consider the rotational asymmetry observed by \citet{Rankin2006}, if real (and assuming their viewing geometry to be correct), to be due to something other than the geometric effects described in this work.

\subsection{Other considerations}

We briefly remark on the difference between using the \citeauthor{Deutsch1955} field and the simpler, but strictly incorrect rotating ``static'' dipole on the numerical results presented in this paper.
In all cases a comparison found that the difference at emission heights $\lesssim 10\%$ of the light cylinder radius was negligible.
A comparison of the predicted emission heights of B0809+74 between the full Deutsch vacuum model and the dipole model is shown in Fig. \ref{fig:emissionheights}.

Finally, our analysis has ignored the possibility of magnetospheric propagation effects \citep[e.g.][]{Barnard1986a}, which may affect the viewing angle of the subpulses (and thus the morphology of the drift bands).
This may ultimately be responsible for the azimuthal asymmetry, but it remains to be shown whether the relatively small height differences involved can account for the large ``pitch angle'' of the skewed beamlets.
The assumption of negligible magnetospheric effects also implies that the polarisation angle is entirely determined by the acceleration vector of the particle as it passes through the visible point, as per Eq. \eqref{eqn:acceleration}.
In that case, the spread of the polarisation angles observed at any given phase (e.g. the histogram in the bottom panel of Fig. \ref{fig:pa}), if significant, is a direct measure of the range of heights that are sampled, and hence the range of $s$ values, providing another means of constraining the emission geometry.
This will be explored more thoroughly in future applications of the geometric model to pulsars with measured polarisation angle histograms.

\section{Conclusion}
\label{sec:conclusion}

We have described a geometric model for pulsar emission that assumes coherent curvature radiation from a corotating, relativistic plasma, consistent with AR effects.
The further assumption (implicit in the carousel model) that the ``activity'' along field lines is only dependent on the spark activity at their footpoints near the stellar surface enables the model to predict the morphology of drift bands without reference to any particular coherent emission mechanism.

We have explored two specific pulsars, PSRs B0809+74 and B2034+19, with contrasting viewing geometries (nearly-aligned and oblique rotators, respectively) and shown that the geometric model can reproduce the essential characteristics of the observed pulsestacks, including the modulation periodicity $P_2$, and drift band curvature.
In the case of B0809+74, we have shown how the geometric model is capable of generating rotational asymmetries in the polar maps produced by the cartographic transform, qualitatively similar to those observed by \citet{Rankin2006}.
However, in this particular case, the directional sense of the asymmetries and the emission height range needed to reproduce them in simulation are difficult to reconcile with their interpretation in the current geometric context.

The geometric model presented in this paper is, we believe, a necessary consequence of the assumptions of curvature radiation, and is therefore applicable to all radio pulsars for which some form of coherent curvature radiation is believed to be the primary emission mechanism, including but not limited to the class of subpulse drifters.
The most telling tests of the validity of this model require high time resolution, broadband observations of individual pulses, especially where microstructure can be resolved---a kind of observation that is scarce, due to the difficulty of coordinating multiple telescopes for a wide spectral coverage.
However, with several broadband instruments coming online (e.g. the ultra-wideband receiver at Parkes, the upgraded GMRT, and the RRI-GBT multiband receiver; \citealt{Maan2013}), in-depth investigations of these subtle effects will become possible in the near future.

\appendix

\section{Coordinate systems}
\label{sec:coordsys}

The derivation of Eq. \eqref{eqn:aligned_Gamma_approx} and, in general, any Taylor expansion about the magnetic axis that depends on Eqs. \eqref{eqn:velocity} and \eqref{eqn:acceleration}, requires the conversion between the \emph{observer} coordination system and the \emph{magnetic} coordinate system.

The observer coordinate system is defined as follows.
The origin is placed at the centre of the pulsar.
The pulsar's rotation axis is identified with the $z$-axis.
The $x$-axis is chosen to lie in the plane spanned by the $z$-axis and the line of sight, $\hat{\bm{n}}$, such that positive $x$ points towards the distant observer (i.e. $\hat{\bm{n}}\cdot\hat{\bm{x}} \ge 0$).
Finally, $\hat{\bm{y}} = \hat{\bm{z}}\times\hat{\bm{x}}$.

The magnetic coordinate system is related to the observer coordinate system by a rotation about the $y$-axis by the magnetic inclination angle $\alpha$, followed by a rotation about the $z$-axis by the rotation phase angle $\phi$.
This brings the $z$-axis into alignment with the magnetic axis, $\hat{\bm{\mu}}$.
When $\phi = 0$, the magnetic axis is in the plane spanned by the $x$- and $z$-axes.
We note that although the magnetic coordinate system does not represent an inertial frame, the magnetic field ($\vec{\bm{B}}$), the velocity field $\vec{\bm{V}}$, and the acceleration field ($\vec{\bm{A}}$) are all static in the co-rotating frame.
That is to say, an inertial observer would find that these fields all corotate with the pulsar, so that a measurement of the entire fields at two different times would be related by a simple rotation about the $z$-axis \citep[see also Appendix A of][]{Dyks2004a}.

Therefore, there is no loss of generality by evaluating these fields at a single rotation phase, say at the fiducial point ($\phi = 0$) and rotating it a posteriori, or equivalently letting the line of sight ``rotate'' in the opposite direction by the same amount.
In many cases, this makes the evaluation of Eq. \eqref{eqn:velocity} and other derived quantities much simpler since they can be expressed independently of $\phi$.
Furthermore, because the only radiation that would be observed is that produced by particles whose velocity vector is parallel to the line of sight, we can simply replace all instances of $\hat{\bm{n}}$ with $\hat{\bm{v}}$ and tacitly ignore times and locations in the magnetosphere where they are not parallel.

The rotation matrices are given many times in the literature, and are not repeated here.
For example, the Appendix of \citet{Yuen2016} includes the relevant transformation matrices, as well as the matrix used to convert between Cartesian and spherical polar coordinates, which is also used in this work to Taylor expand about the magnetic axis (i.e. about $\theta = 0$).

\bibliography{biblio}

\end{document}